\documentclass[twocolumn,showpacs,preprintnumbers,amsmath,amssymb]{revtex4}
\usepackage{graphicx}
\usepackage{dcolumn}
\usepackage{bm}

\begin{document}

\title{The Clausius-Mossotti Phase Transition in Polar Liquids}
\author{S. Sivasubramanian}
\author{A. Widom}
\author{Y.N. Srivastava$^\dagger$}

\affiliation{Physics Department, Northeastern University, Boston MA USA \\
$^\dagger$Physics Department \& INFN, University of Perugia, Perugia Italy}

\begin{abstract}
The conventional Clausius-Mossotti polarization equation of state is
known to be unstable for polar liquids having molecules with high
polarizability. Room temperature water is an important example.
The instability in the polarization equation of state is of the
typical loop form requiring an ``equal area'' construction
for studying the stable ordered phase. The ordered phase of a
Clausius-Mossotti polar liquid then consists of domains each having
a net polarization. The polarization may vary in direction
from domain to domain. The ordered phases are quite similar to those
previously discussed on the basis of Dicke superradiance.
\end{abstract}

\maketitle

\section{Introduction}

The Clausius-Mossotti\cite{Mossotti,Clausius} polarization
equation of state is quite often discussed in text books on
electromagnetic theory\cite{David,Becker,Eyges}. The discussion
is thought to give students a good physical idea about how
microscopic models of dipole moments yield non-trivial macroscopic
material dielectric constants \begin{math} \varepsilon \end{math}.
It is only sometimes mentioned that the Clausius-Mossotti model exhibits an
instability hinting at a phase transition into an ordered polarized state.
The unstable precursor to the phase transition is in fact self evident.
The instability is particularly strong in polar liquids such as water.
In fact, the Clausius-Mossotti model in its classic text book form
indicates that water at room temperature and pressure is in an ordered
polarization state.

The ordered state of polar liquids is described in the literature as
exhibiting {\em superradiance}\cite{Prep1,Del1,Del2,Siva}. In the
superradiant state, the polarization \begin{math} {\bf P} \end{math}
has a mean value roughly constant in magnitude within an ordered domain.
However, \begin{math} {\bf P} \end{math} may vary in direction as one
goes from one ordered fluid domain to another ordered fluid domain.
The theoretical derivation of superradiant domains normally starts
from condensed matter quantum electrodynamics. Yet many insights into
the nature of superradiant  thermal equilibrium can be obtained from
a classical electrodynamic viewpoint. Our purpose is to derive the polar
liquid phase transition implicit in the Clausius-Mossotti model. Our
arguments employ (apart from classical electrodynamics) the elementary
statistical thermodynamics previously worked out by Debye and
Langevin\cite{Debye,Langevin}.

In Sec. II, we briefly discuss the derivation of the Clausius-Mossotti
{\em local} electric field (in Gaussian units)
\begin{equation}
{\bf F}={\bf E}+(4\pi /3){\bf P}.
\label{LocalField}
\end{equation}
In Eq.(\ref{LocalField}), \begin{math} {\bf E} \end{math} is the total
electric field and \begin{math} {\bf P} \end{math} is the dipole moment per
unit volume. In terms of the molecular polarizability
\begin{math} \alpha  \end{math} of a
single molecule and the liquid volume
\begin{math} v \end{math} per molecule, we shall derive
the mean field Clausius-Mossotti dielectric constant
\begin{math} \varepsilon  \end{math}.
In Sec. III, we discuss the region of thermodynamic stability of the
Clausius-Mossotti model. The known {\em instability} is often thought
to be an {\em embarrassment} for the mean field theory.
However, we argue that the Clausius-Mossotti instability is a signature of a
physical phase transition into an ordered phase. The ordered phase has been
previously discovered from the viewpoint of superradiance. The thermodynamic
theories of both the ordered and disordered phases are discussed from
the Clausius-Mossotti viewpoint in Sec. IV. The nature of a more complete
quantum theory is explored in the concluding Sec. V.

\section{Clausius-Mossotti Local Fields}

Consider the Coulomb energy contained in a charge density
\begin{math} \rho \end{math}, i.e.
\begin{equation}
U=\frac{1}{2}\int \int \frac{\rho({\bf r})\rho({\bf r}^\prime )}
{|{\bf r}-{\bf r}^\prime |} \ d^3{\bf r}d^3{\bf r}^\prime .
\label{LocalField1}
\end{equation}
If the charge density is derived from a macroscopic polarization
\begin{equation}
\rho({\bf r})=-div{\bf P}({\bf r}),
\label{LocalField2}
\end{equation}
then integrating Eq.(\ref{LocalField1}) twice by parts yields
\begin{equation}
U=\frac{1}{2}\int \int {\bf P}({\bf r})\cdot {\sf G}({\bf r}-{\bf r}^\prime )
\cdot {\bf P}({\bf r}^\prime)\ d^3{\bf r}d^3{\bf r}^\prime .
\label{LocalField3}
\end{equation}
The dyadic
\begin{equation}
{\sf G}({\bf r}-{\bf r}^\prime )={\bf grad}\ {\bf grad}^\prime \
\frac{1}{|{\bf r}-{\bf r}^\prime |}
\label{LocalField4}
\end{equation}
obeys the trace condition
\begin{equation}
Tr{\sf G}({\bf r}-{\bf r}^\prime )=-\nabla^2
\ \frac{1}{|{\bf r}-{\bf r}^\prime |}=4\pi \delta ({\bf r}-{\bf r}^\prime ).
\label{LocalField5}
\end{equation}
Eqs.(\ref{LocalField4}) and (\ref{LocalField5}) imply that
\begin{eqnarray}
{\sf G}({\bf R})&=&-{\sf \Delta }({\bf R})+
\frac{4\pi}{3} {\sf 1}\delta ({\bf R}), \nonumber \\
{\sf \Delta }({\bf R})&=&
\frac{3{\bf RR}-R^2 {\sf 1}}{R^5}\ ,
\label{LocalField6}
\end{eqnarray}
where the \begin{math} (4\pi {\sf 1}/3)\delta ({\bf R}) \end{math}
term in Eq.(\ref{LocalField6}) is present to enforce the
trace condition in Eq.(\ref{LocalField5}).
From Eqs.(\ref{LocalField3}) and (\ref{LocalField6}), it follows
that the Coulomb energy
\begin{equation}
U=U_{dipole}+U_{local},
\label{LocalField7}
\end{equation}
where
\begin{equation}
U_{dipole}=
-\frac{1}{2}\int \int {\bf P}({\bf r})\cdot
{\sf \Delta }({\bf r}-{\bf r}^\prime )
\cdot {\bf P}({\bf r}^\prime)\ d^3{\bf r}d^3{\bf r}^\prime
\label{LocalField8}
\end{equation}
and
\begin{equation}
U_{local}=\frac{2\pi }{3}\int |{\bf P}({\bf r})|^2 d^3{\bf r}.
\label{LocalField9}
\end{equation}

Consider the work done by bringing in from infinity a small change
in the charge density
\begin{math}
\delta \rho =-div \delta {\bf P}
\end{math}.
The change in Coulomb energy will be given in terms of the potential
\begin{math} \Phi({\bf r}) \end{math} according to
\begin{eqnarray}
\delta U &=& \int \int
\frac{\rho({\bf r}^\prime )\delta \rho({\bf r})}{|{\bf r}-{\bf r}^\prime |}
d^3{\bf r}^\prime d^3{\bf r}
\nonumber \\
&=& \int \Phi({\bf r})\delta \rho({\bf r})d^3{\bf r}
\nonumber \\
&=& -\int \Phi({\bf r})div \delta {\bf P}({\bf r})d^3{\bf r}
\nonumber \\
&=& \int {\bf grad}\Phi({\bf r})\cdot \delta {\bf P}({\bf r})d^3{\bf r}.
\label{LocalField10}
\end{eqnarray}
The electric field \begin{math} {\bf E}=-{\bf grad}\Phi \end{math} may
then be computed from the functional variation
\begin{eqnarray}
\delta U &=& -\int {\bf E}({\bf r})\cdot
\delta {\bf P}({\bf r})d^3{\bf r}, \nonumber \\
{\bf E}({\bf r}) &=&
{\bf F}({\bf r})-\frac{4\pi}{3}{\bf P}({\bf r}).
\label{LocalField11}
\end{eqnarray}
The local electric field
\begin{math} {\bf F}({\bf r})  \end{math}
is due directly to the dipole-dipole interaction
function \begin{math} {\sf \Delta }({\bf R})  \end{math}
in Eq.(\ref{LocalField6}) via the variation
\begin{eqnarray}
\delta U_{dipole} &=& -\int {\bf F}({\bf r})\cdot
\delta {\bf P}({\bf r})d^3{\bf r} , \nonumber \\
{\bf F}({\bf r})&=& \int {\sf \Delta}({\bf r}-{\bf r}^\prime )
\cdot {\bf P}({\bf r}^\prime )d^3{\bf r}^\prime .
\label{LocalField12}
\end{eqnarray}
The remaining \begin{math} (4\pi /3){\bf P}({\bf r}) \end{math}
term on the right hand side of Eq.(\ref{LocalField11}) is due to the
local polarization energy in  Eq.(\ref{LocalField9}), i.e.
\begin{eqnarray}
\delta U_{local} &=& \delta U-\delta U_{dipole}
\nonumber \\
&=& -\int \{ {\bf E}({\bf r})-{\bf F}({\bf r}) \}
\cdot \delta {\bf P}({\bf r})d^3{\bf r}
\nonumber \\
&=&
\frac{4\pi }{3}\int {\bf P}({\bf r})
\cdot \delta {\bf P}({\bf r})d^3{\bf r}.
\label{LocalField13}
\end{eqnarray}
Eq.(\ref{LocalField13}) implies the Clausius-Mossotti
local field Eq.(\ref{LocalField}).

Now suppose that the individual fluid molecules exhibit a mean electric
dipole moment response to a local field as described by the polarizability
\begin{math} \alpha  \end{math} via
\begin{equation}
\bar{\bf p}=\alpha {\bf F}.
\label{LocalField14}
\end{equation}
The dipole moment per unit volume (corresponding to a volume
\begin{math} v \end{math} per molecule of fluid) then reads
\begin{equation}
{\bf P}=\frac{\bar{\bf p}}{v}=\frac{\alpha {\bf F}}{v}=
\frac{\alpha }{v}\left({\bf E}+\frac{4\pi }{3}{\bf P}\right).
\label{LocalField15}
\end{equation}
Thus
\begin{equation}
{\bf P}=\chi {\bf E},
\label{LocalField16}
\end{equation}
with a polarization susceptibility
\begin{equation}
\chi =\left(\frac{\alpha }{v-(4\pi /3)\alpha}\right).
\label{LocalField17}
\end{equation}
In terms of the dielectric constant
\begin{math} \varepsilon  \end{math},
\begin{equation}
{\bf D}={\bf E}+4\pi {\bf P}=\varepsilon {\bf E},
\label{LocalField18}
\end{equation}
one finds the Clausius-Mossotti prediction
\begin{equation}
\varepsilon =1+4\pi \chi
=\left(\frac{v+(8\pi /3)\alpha }{v-(4\pi /3)\alpha}\right).
\label{LocalField19}
\end{equation}
All that is needed to calculate
\begin{math} \varepsilon  \end{math}
(in this model) is the polarizability
\begin{math} \alpha  \end{math} for
an isolated molecule and the volume
per molecule \begin{math} v \end{math}
in the fluid.

\section{Thermodynamic Stability}

Landau and Lifshitz\cite{Landau1} prove a non-trivial theorem
in their treatise on the electrodynamics of continuous media.
The theorem asserts that the dielectric constant of a material
obeys the inequality
\begin{equation}
1<\varepsilon <\infty \ \ \ {\rm (thermodynamic\ stability)}.
\label{ThermalStability1}
\end{equation}
We note (in passing) that the magnetic analogue of the theorem
(with \begin{math} {\bf B}=\tilde{\mu }{\bf H} \end{math}) asserts
only the weaker magnetic permeability inequality
\begin{math}0<\tilde{\mu }<\infty  \end{math}. Thus, diamagnetism
(\begin{math} 0<\tilde{\mu }<1  \end{math}) can exist but diaelectricity
(\begin{math} 0<\varepsilon <1  \end{math}) cannot exist.
If one applies the thermodynamic stability criteria of the
Landau-Lifshitz Eq.(\ref{ThermalStability1}) to the Clausius-Mossotti
Eq.(\ref{LocalField19}), then it becomes apparent that the region of
stability for the model may be described by the parameter
\begin{eqnarray}
\eta &=& \left(\frac{4\pi \alpha }{3v}\right),
\nonumber \\
\eta &<& 1 \ \ {\rm (thermodynamic\ stability)}.
\label{ThermalStability2}
\end{eqnarray}

Debye defines a {\em polar molecule} as one whose polarizability
has a temperature dependence (in the dilute gas phase) of the form
\begin{equation}
\alpha =\alpha_0+\left(\frac{\mu^2}{3k_BT}\right).
\label{Polar1}
\end{equation}
By plotting \begin{math} \alpha \end{math} as a function of
inverse temperature one finds a linear slope yielding the magnitude
of a ``permanent'' dipole moment
\begin{equation}
\mu =|{\bf p}|.
\label{Polar2}
\end{equation}
If the polarizability of a single molecule (say in the gas phase)
exhibits an appreciable permanent dipole moment
\begin{math} \mu  \end{math}, then the molecule is called polar.
The water molecule \begin{math} H_2O \end{math} is an important
example of a polar molecule. In the dilute gas phase, a water molecule
exhibits a temperature dependent polarizability of the form in
Eq.(\ref{Polar1}) with
\begin{eqnarray}
\alpha_0(H_2O) &\approx & 1.494\times 10^{-24}\ cm^3,
\nonumber \\
\mu(H_2O) &\approx & 1.855 \times 10^{-18}\ Gauss\ cm^3.
\label{Polar3}
\end{eqnarray}
For room temperature and atmospheric pressure, well known properties of
water\cite{CRC} include
\begin{eqnarray}
\alpha_{water} &\approx & 28.696\ \AA^3\ \ ({\rm Gas\ Phase}),
\nonumber \\
v_{water} &\approx & 30.014\ \AA^3\ \ ({\rm Liquid\ Phase}),
\label{Polar4}
\end{eqnarray}
and
\begin{equation}
\eta_{water}=\left(\frac{4\pi \alpha_{water}}{3v_{water}}\right)
\approx 4.0054 >1\ \ {\rm (unstable)}.
\label{Polar5}
\end{equation}
In TABLE {\ref{table1}} we have listed some liquids
along with a computation of whether or not the conventional
Clausius-Mossotti model yields a stable polarization
{\em disordered} state.

\begin{table}[bp]
\caption{\label{table1} Listed below are some liquids\cite{CRC}.
The stability properties are evaluated on the basis of the Clausius-Mossotti
model Eq.(\ref{ThermalStability2}) with $\eta =(4\pi \alpha /3v)$.}
\begin{ruledtabular}
\begin{tabular}{lcl}
Chemical & $\eta $ & Clausius-Mossotti\\
\hline \\
 $C_2H_5OC_2H_5$ & 0.5486 & stable \\
 $HCONH_2$ & 6.1498 & unstable \\
 $HCN$ & 4.7479 & unstable \\
 $HF$ & 3.6503 & unstable \\
 $HCONHCH_3$ & 5.2093 & unstable \\
 $C_6H_5CH_3$ & 0.3156 & stable \\
 $PCl_3$  & 0.6728 & stable \\
 $H_2O$  & 4.0054 & unstable \\
 $ CHCl_3$ & 0.5538 & stable \\
\end{tabular}
\end{ruledtabular}
\end{table}
Room temperature (and pressure) water is an important
example of a substance which is in an unstable regime
of the Clausius-Mossotti model. Many other polar
liquids also tend to be in unstable regimes. In their studies
of polar liquids, the well known physical chemists Onsager\cite{Onsager}
and (later) Kirkwood\cite{Kirkwood,Frohlich} tried to improve on the
Clausius-Mossotti theory in such a manner that the instability
would be avoided. The attempts were not entirely successful.

More recently, there have been studies which assert that polar liquids
(such as water) may be found in a ``superradiant'' state wherein ordered
domains exist. We regard these more recent proposals as likely to be true.
The superradiant ordered sub-domains of the liquid carry a net polarization
\begin{math} {\bf P}_0 \end{math}. From the viewpoint of Clausius-Mossotti
local fields, the polarized sub-domains should in reality be present.
In deriving this result below, we shall neglect
\begin{math} \alpha_0 \end{math} in Eq.(\ref{Polar1}) compared
with the thermal term, i.e. we employ the Langevin model form
\begin{equation}
\alpha_T = \left(\frac{\mu^2}{3k_BT}\right)
\ \ {\rm (Langevin)}.
\label{Polar6}
\end{equation}
This is certainly a good approximation for water wherein
\begin{math} \alpha_{0}(H_2O)<<\alpha_{water}\approx \alpha_T \end{math}.
In what follows we shall also review how Eq.(\ref{Polar6}) is derived.

\section{Ordered Thermodynamic Phases}

For a permanent dipole moment
\begin{math}{\bf p}=\mu {\bf n} \end{math}
(where \begin{math}{\bf n} \end{math}
is a unit vector), the partition function describing
the interaction with the local field
\begin{math} {\bf F} \end{math} is given by
\begin{equation}
Z=\int e^{{\bf p\cdot F}/k_BT}\left(\frac{d^2{\bf n}}{4\pi }\right).
\label{Ordered1}
\end{equation}
In Eq.({\ref{Ordered1}}), \begin{math} d^2{\bf n} \end{math} is a solid
angle about the \begin{math} {\bf n} \end{math} direction. Explicitly,
\begin{equation}
Z=\frac{1}{2}\int_0^\pi e^{\mu F\cos\theta /k_BT}sin\theta d\theta ,
\label{Ordered2}
\end{equation}
yielding a free energy per unit volume of
\begin{equation}
A=-\frac{k_BT}{v}\ln Z
=-\frac{k_BT}{v}\ln \left[\frac{\sinh(\mu F/k_BT)}{(\mu F/k_BT)}\right].
\label{Ordered3}
\end{equation}
The resulting polarization
\begin{math} {\bf P}=-(\partial A/\partial{\bf F})_T \end{math}
is parallel to \begin{math} {\bf F} \end{math} and has the magnitude
originally derived by Langevin, i.e.
\begin{equation}
P=\frac{\mu }{v}\left[coth\left(\frac{\mu F}{k_BT}\right)
-\left(\frac{k_BT}{\mu F}\right)\right].
\label{Ordered4}
\end{equation}
Note that
\begin{equation}
\frac{\alpha_T}{v}=\lim_{F\to 0}
\left(\frac{\partial P}{\partial F}\right)_T
=\left(\frac{\mu^2}{3vk_BT}\right),
\label{Ordered5}
\end{equation}
in agreement with Eq.(\ref{Polar6}).

From Eqs.(\ref{LocalField}) and (\ref{Ordered4}) we may compute
(in parametric form) the polarization equation of state as
follows:
\par \noindent
(i) Define the parameters
\begin{eqnarray}
\eta &=& \frac{4\pi \alpha_T}{3v},
\ \ \ x = \frac{\mu E}{k_BT}\ ,
\nonumber \\
y &=& \frac{\mu P}{\eta k_BT}\ ,
\ \ \ z = \frac{\mu F}{k_BT}\ .
\label{Ordered6}
\end{eqnarray}
(ii) Eqs.(\ref{LocalField}) and (\ref{Ordered4})
now read
\begin{eqnarray}
x &=& z-3\eta \left[\coth(z)-\frac{1}{z}\right],
\nonumber \\
y &=& \frac{9}{4\pi}\left[\coth(z)-\frac{1}{z}\right].
\label{Ordered7}
\end{eqnarray}
(iii) By eliminating the parameter \begin{math} z \end{math}
in Eqs.(\ref{Ordered7}) one defines the function
\begin{equation}
y={\cal F}(x,\eta )
\label{Ordered8}
\end{equation}
and thereby the non-linear polarization equation of state
\begin{equation}
P=\frac{4\pi \mu}{9v }
{\cal F}\left(\frac{\mu E}{k_BT}\ ,\frac{4\pi \mu^2}{9vk_BT}\right).
\label{Ordered9}
\end{equation}
The polarization is plotted in FIG. 1 for room temperature toluene
(which is disordered) and water (which is ordered).
\begin{figure}[bp]
\scalebox {0.4}{\includegraphics{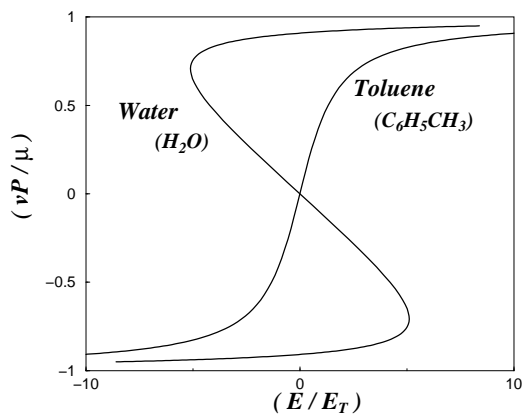}}
\caption{The Clausius-Mossotti ``dimensionless'' polarization
$(vP/\mu)$ is plotted as a function
of the electric field, in units of the thermal electric field
$E_T=(k_BT/\mu)$. Toluene is stable with a positive susceptibility
$\chi_T=(\partial P/\partial E)_T$. Water is unstable as indicated
by the unphysical ``$\chi_T<0$'' portion of the
loop in the equation of state. The stable phase of water may
be constructed using an ``equal area'' rule.}
\label{fig1}
\end{figure}

The polarization equation of state in the Langevin-Clausius-Mossotti
model is of the typical ``mean field'' variety. There exists a critical
temperature
\begin{equation}
T_c=\left(\frac{4\pi \mu^2}{9k_Bv}\right) .
\label{Ordered10}
\end{equation}
Above the critical temperature, the susceptibility is positive,
\begin{equation}
\left(\frac{\varepsilon_T-1}{4\pi }\right)=
\chi_T=\left(\frac{\partial P}{\partial E}\right)_T > 0
\ \ {\rm for}\ \ T>T_c ,
\label{Ordered11}
\end{equation}
in accordance with the Landau-Lifshitz Eq.(\ref{ThermalStability1}).
Below the critical temperature, there is an unphysical
``\begin{math}\chi_T<0\end{math}'' portion of an equation
of state loop which must be subject to an ``equal area''
construction\cite{Ferroequal} to deduce
the proper stable phase. There exists below the critical
temperature a remnant polarization
\begin{equation}
P_0(T)=\lim_{E\to 0^+} P(E,T)>0
\ \ {\rm for}\ \ T<T_c .
\label{Ordered12}
\end{equation}
The parametric equations for the remnant polarization are given by
\begin{eqnarray}
\frac{T}{T_c}=\frac{3}{z}\left(coth(z)-\frac{1}{z}\right),
\nonumber \\
\frac{vP_0}{\mu}=\left(coth(z)-\frac{1}{z}\right),
\label{Ordered13}
\end{eqnarray}
where the parameter range is \begin{math}0<z<\infty \end{math}.
The ordered phase of the Langevin-Clausius-Mossotti polar liquid is
ferroelectric with a polarization pointing in a unit vector
direction \begin{math} {\bf n}=({\bf P}_0/P_0) \end{math}.
The magnitude of the remnant polarization is plotted in FIG. 2.

\begin{figure}[tp]
\scalebox {0.4}{\includegraphics{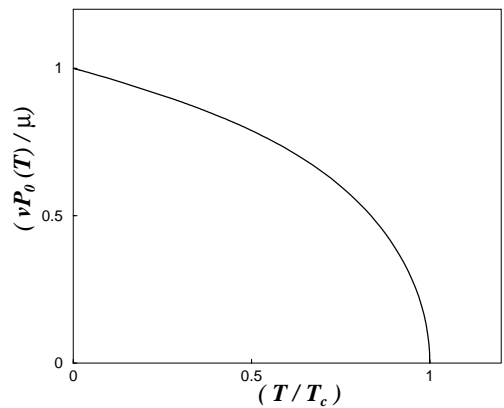}}
\caption{Shown is the remnant polarization $P_0(T)$ for
temperatures below the critical temperature. The ordered phase of
the Langevin-Clausius-Mossotti polar liquid is ferroelectric. The
ordered net polarized regions should appear in mesoscopic domains.
The direction of the remnant polarization
${\bf n}=({\bf P}_0/P_0)$
should vary depending upon the domain.}
\label{fig2}
\end{figure}

In this picture, water at room temperature contains ordered
ferroelectric domains with a polarization pointing in the direction
\begin{math} {\bf n} \end{math}. From one domain to another, the
direction \begin{math} {\bf n} \end{math} of the polarization is changing.
The situation is shown schematically in FIG. 3. Inside of a domain,
there is a residual local electric field
\begin{math} {\bf F}_0 \end{math} and polarization
\begin{math} {\bf P}_0 \end{math} related by
\begin{equation}
{\bf F}_0=\left(\frac{4\pi }{3}\right){\bf P}_0 .
\label{Ordered14}
\end{equation}

\begin{figure}[tp]
\scalebox {0.35}{\includegraphics{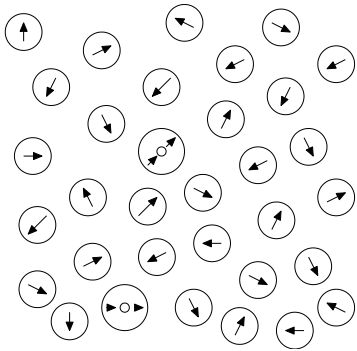}}
\caption{Shown {\it schematically} is a system of polarized domains
in water. In each domain there exists a local electric field and a
polarization vector related by ${\bf F}_0=(4\pi /3){\bf P}_0$. Between
the ordered phase domains are unpolarized disordered water. Two of the
domains are shown to contain a noble gas atom (such as Argon). The low
entropy ordered domain about a noble gas impurity has been called
an ``iceberg'' and was unexpected when it was first measured
thermodynamically.}
\label{fig3}
\end{figure}

The domain picture has many interesting features only one of which we
shall discuss here. If one asks what will happen to a noble gas impurity
(say Argon) when it is dissolved in water, then one concludes that the
atom will be drawn inside the ordered domain. The effective
potential\cite{Landau2} on the atom is (roughly) a step function
attracting the impurity into the ordered domain,
\begin{equation}
\Delta U_{impurity}({\bf r})=
-\frac{1}{2}\alpha_{impurity}|{\bf F}_0({\bf r})|^2 .
\label{Ordered15}
\end{equation}
In FIG. 3 we show two of the domains with an impurity. The notion that
an ordered domain is clustered around a dissolved
noble gas atom (say argon) has consequences for the thermodynamic impurity
equations of state. The impurity entropy has in fact been particularly well
analyzed by Frank and Evans\cite{Iceberg1,Iceberg2}. Suppose that a dilute
mixture of Argon in water is such that the vapor phase mixture is
in thermal equilibrium with the dilute liquid mixture. An impurity atom
in water vapor is in a disordered environment. The impurity atom in
liquid water will be found in an ordered environment within a domain.
Impurity entropy is then {\em lost} when the impurity atom moves from
the vapor to the liquid. To use the words of Frank and Evans
(describing this unexpected entropy loss) ``the water builds a microscopic
iceberg around the non-polar molecule''. There is no evidence that an ordered
domain about a noble gas atom impurity is a piece of ice. The low entropy
polarized domain predicted by the Langevin-Clausius-Mossotti model appears
to explain the so called ``iceberg effect''.

\section{Conclusions}

The Clausius-Mossotti model together with the elementary
statistical thermodynamics of Debye and Langevin yields a
ferroelectric phase transition for many polar liquids including
water. The notion of ferroelectric domains within the a polar
liquid with a net polarization
\begin{math} {\bf P}_0 \end{math} and internal electric field
\begin{math} {\bf F}_0=(4\pi /3){\bf P}_0 \end{math} is clearly
implicit in the model and can be derived above in a fairly
simple and straight forward manner. The phase transition may be
described using basic concepts from elementary physics.

Nevertheless, reasoning with classical electrodynamics
and classical statistical thermodynamics yields only partial
insights concerning the complete microscopic theory. A detailed
theory involves quantum statistical mechanics and quantum
electrodynamics. For example, why does a
polar \begin{math} H_2O \end{math} molecule have a permanent
electric dipole moment? If one starts by calculating the quantum
ground state \begin{math} \left|0\right> \end{math}
of the molecule employing only the non-relativistic Coulomb
Hamiltonian, then parity conservation requires the vanishing of the
mean dipole moment
\begin{math}
\bar{\bf p}=\left<0\left|{\bf p}\right|0\right>=0
\end{math}. The notion of a permanent electric
dipole moment requires a break in parity (space inversion) symmetry.
Such a symmetry break can come about due to interactions with the
quantum electrodynamic field. Similarly, the dipole-dipole interaction
takes place (in the quantum electrodynamic theory) due to the
exchange of virtual photons between two dipoles.

That the polarized domains have a net internal electric
field \begin{math} {\bf F}_0 \end{math}, really means that the
exchanged photons form a Bose condensate. The Bose condensate is
also present in a model conventionally used to describe
superradiance\cite{HeppLieb}. In spite of the long time popularity
of the Clausius-Mossotti model, it was only through a study of superradiance
that the ordered domains were first\cite{OriginWater} theoretically
discovered. It is hoped that the above simple Clausius-Mossotti model
considerations serve to clarify the mechanism for ordering in some
polar liquids.


\begin{thebibliography}{03}
\bibitem{Mossotti} O.F. Mossotti, {\it Memorie Mat. Fis. Modena} {\bf 24},
49 (1850).
\bibitem{Clausius} R. Clausius, {\it Die mechanische W. armtheorie} {\bf II},
62 Braunschweig (1897).
\bibitem{David} D.J. Griffiths, ``Introduction to Electrodynamics'',
pp. 192 (Prentice Hall, New Jersey 1989).
\bibitem{Becker} R. Becker, ``Electromagentic Fields and Interactions'',
pp. 95 (Dover, New York 1982).
\bibitem{Eyges} L. Eyges, ``The Classical Electromagnetic Field'',
pp. 111 (Dover, New York 1972).
\bibitem{Prep1} G. Preparata, {\it QED Coherence in Matter}
chap. 10, pp. 195 (World Scientific, Singapore, 1995).
\bibitem{Del1} E. Del Giudice and G. Preparata,
{\it A New QED Picture of Water in Macroscopic Quantum Coherence}
eds. E. Sassaroli, Y. Srivastava, J. Swain and A. Widom
(World Scientific, Singapore, 1998).
\bibitem{Del2} E. Del Guidice, G. Preparata and M. Fleischmann,
{\it J. Elec. Chem.} {\bf 482}, 110 (2000).
\bibitem{Siva} S. Sivasubramanian, A. Widom and Y.N. Srivastava,
{\it Mod. Phys. Lett.} {\bf B16}, 1201 (2002).
\bibitem{Debye} P. Debye, {\it Phys. Z.} {\bf 13}, 97 (1912);
``Polar Molecules'', pp. 30 (Dover Publications Inc. New York. 1928).
\bibitem{Langevin} P. Langevin, {J. Phys.} {\bf 4}, 678 (1905);
{\it Ann. chim. phys.} {\bf 5}, 70 (1905).
\bibitem{Landau1} L.D. Landau, E.M. Lifshitz, ``Electrodynamics of
Continuous Media'', chap. II pp. 55 (Pergamon Press, New York 1975).
\bibitem{CRC} ``CRC Handbook of Chemistry and Physics'',
(CRC Press, Cleveland, OH, 2001).
\bibitem{Onsager} L. Onsager, {\it J. Amer. Chem. Soc.} {\bf 58}, 1486 (1936).
\bibitem{Kirkwood} J.G. Kirkwood, {\it J. chem. Phys.} {\bf 7}, 911 (1939);
G. Oster and J.G. Kirkwood, {\it J. chem. Phys.} {\bf 11}, 175 (1943);
J.G. Kirkwood, {\it Trans. Faraday Soc.} {\bf A42}, 7 (1946).
\bibitem{Frohlich} H. Fr\"ohlich, ``Theory of Dielectrics'',
Oxford University Press (1949).
\bibitem{Ferroequal} B.A. Strukov and A.P. Levanyuk, ``Ferroelectric
Phenomena in Crystals'', pp. 57 (Springer, New York 1997).
\bibitem{Landau2} L.D. Landau, E.M. Lifshitz, ``Quantum Mechanics'',
chap. XI pp. 341 (Pergamon Press, New York 1981).
\bibitem{Iceberg1} H.S. Frank and M.W. Evans, {\it J. Chem. Phys.}
{\bf 13}, 507 (1945)
\bibitem{Iceberg2} R.A. Robinson and R.H. Stokes,
``Electrolyte Solution'',  Second Edition chap. I pp. 14
(Dover, New York 2002).
\bibitem{HeppLieb} K. Hepp and E.H. Lieb, {\it Phys. Rev.} {\bf A8}, 2517 (1973);
{\it Ann. Phys.} {\bf 76}, 360 (1973).
\bibitem{OriginWater} E. Del Giudice, G. Preparata and G. Vitiello,
{\it Phys. Rev. Lett.} {\bf 61}, 1085 (1988)
\end{thebibliography}
\end{document}